\documentclass[aip, reprint]{revtex4-2}
\usepackage{amsfonts,amsmath,bm,amssymb,color}
\usepackage{graphicx}
\usepackage{hyperref}

\begin{document}

\title{Statistical analysis of vortex condensate motion in two-dimensional turbulence}

\author{Vladimir Parfenyev}\email{parfenius@gmail.com}
\affiliation{Landau Institute for Theoretical Physics, Russian Academy of Sciences, 1-A Akademika Semenova av., 142432 Chernogolovka, Russia}
\affiliation{HSE University, Faculty of Physics, Myasnitskaya 20, 101000 Moscow, Russia}

\begin{abstract}
An inverse turbulent cascade in a periodic square box produces a coherent system-sized vortex dipole. We study the statistics of its motion by carrying out direct numerical simulations performed for various bottom friction $\alpha$, pumping intensity $\varepsilon$, and fluid hyperviscosity $\nu$. In the main approximation, coherent vortices can be considered as point vortices, and within this model, they drift at the same dipole velocity, which is determined by their circulation and mutual arrangement. The characteristic value of the dipole velocity is more than an order of magnitude smaller than the polar velocity inside coherent vortices. Turbulent fluctuations give rise to a relative velocity between the vortices, which changes the distance between them. We found that for a strong condensate, the probability density function of the vector $\bm \rho$, describing the difference in the mutual arrangement of coherent vortices from half the diagonal of the computational domain, has the form of a ring. The radius of the ring weakly depends on control parameters and the width of the ring is proportional to the dimensionless parameter $\delta = \epsilon^{-1/3} L^{2/3} \alpha$, where $\epsilon$ is the inverse energy flux and $L$ is the system size. The random walk around the ring, caused by turbulent fluctuations, has superdiffusion behavior at intermediate times. It results in a finite correlation time of the dipole velocity, which turns out to be of the order of turnover time $\tau_K = L^{2/3} \epsilon^{-1/3}$ of system-size eddies produced by an inverse turbulent cascade. The results obtained deepen the understanding of the processes governing the motion of coherent vortices.
\end{abstract}

\maketitle

\section{Introduction}

The main feature of two-dimensional turbulence is the inverse energy cascade, i.e. the transfer of energy from a relatively small forcing scale to larger scales~\cite{kraichnan1967inertial, leith1968diffusion, batchelor1969computation}. In a square domain with low large-scale friction, the energy piles up at the system size in the form of coherent vortices~\cite{chertkov2007dynamics, xia2009spectrally, chan2012dynamics, bardoczi2014experimental, laurie2014universal, frishman2017jets, parfenyev2022profile, scott2023annular}. The observed vortices have well-defined isotropic mean profiles, which can be characterized by the polar velocity $U_{\varphi}(r)$, where $r$ is the distance to the vortex axis. The mean velocity profiles can be determined analytically if the self-action of turbulent pulsations is small compared to the action of the coherent flow. When large-scale dissipation of energy is determined by the linear bottom friction $\alpha$, then the velocity profile is $r$-independent, $U_{\varphi} = \sqrt{3 \epsilon/\alpha}$, where $\epsilon$ is the inverse energy flux~\cite{laurie2014universal, kolokolov2016structure, frishman2017culmination}. In the absence of bottom friction, the energy dissipation occurs due to the fluid viscosity $\nu$, and the velocity profile is equal to $U_{\varphi} = \sqrt{\epsilon/\nu} \, r \ln (R/r)$, where $R$ is the vortex size~\cite{doludenko2021coherent}. Note that similar predictions are also valid for three-dimensional columnar vortices in rapidly rotating fluids~\cite{kolokolov2020structure, parfenyev2021influence, parfenyev2021velocity}.

In addition to studying the mean velocity profile, much attention was paid to the statistical analysis of turbulent fluctuations against its background in the reference frame associated with the coherent vortex~\cite{kolokolov2016velocity, frishman2018turbulence, kolokolov2023pair}. However, the motion of coherent vortices as a whole remains practically unexplored. It essentially depends on the boundary conditions. If the boundaries are no-slip walls, then the coherent vortex is localized near the center of the square domain~\cite{xia2009spectrally, bardoczi2014experimental, doludenko2021coherent, molenaar2004angular}, and for periodic boundaries, a vortex dipole is formed that moves throughout the entire region~\cite{chertkov2007dynamics, chan2012dynamics, laurie2014universal, parfenyev2022profile}. To the best of our knowledge, the motion of the vortex condensate was systematically investigated only for periodic boundary conditions and in the absence of bottom friction, see Ref.~\onlinecite{chan2012dynamics}. Here we extend the analysis to the case when the large-scale dissipation of energy in a system with periodic boundaries is determined by linear bottom friction. 

\begin{figure*}[t]
\centering{\includegraphics[width=\linewidth]{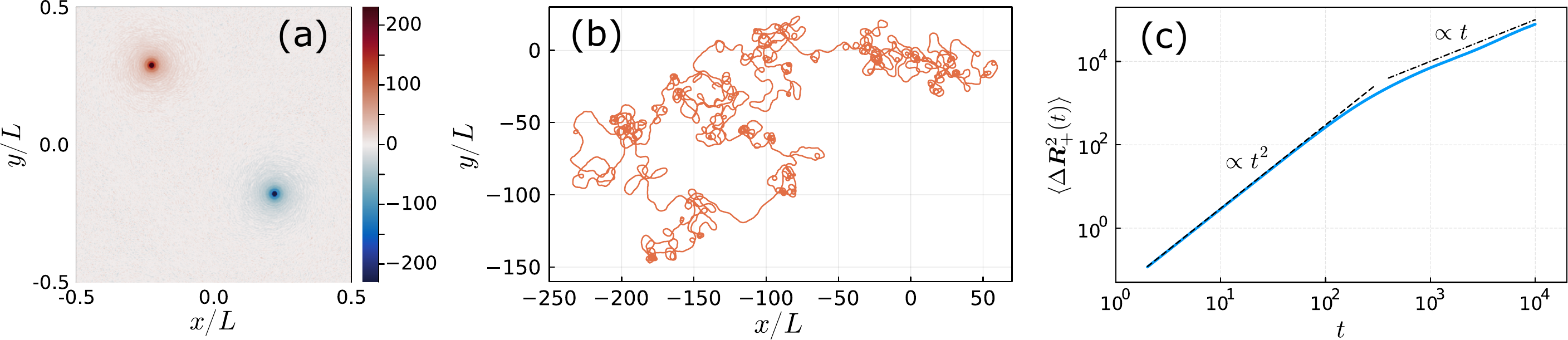}}
\caption{(a) Snapshot of the vorticity field ($512^2$) normalized by $(\epsilon/\alpha L^2)^{1/2}$ in the statistical steady-state. (b) Unfolded trajectory of the positive vortex. (c) The displacement squared of the positive vortex averaged over the unfolded trajectory. The data are given for DNS run $B$, but are typical for all simulations.}
\label{fig:1}
\end{figure*}

In agreement with previous studies, we found that the main energy of the system is concentrated in two coherent vortices of different signs, forming a vortex dipole moving randomly. The radii of the vortices are small compared to the distance between them, and in the main approximation, their influence on each other can be described within the analytical model of point vortices. If there were no turbulent pulsations, then the vortex dipole would move at a constant velocity, determined by the time-invariant mutual arrangement of the vortices. Turbulent fluctuations lead to a relative velocity between the vortices, which changes the distance between them and makes the velocity of the vortex dipole time-dependent. The analytical model of point vortices allows us to determine the dipole velocity at each moment (which is associated with the direct influence of coherent vortices on each other) and to separate it from the motion of coherent vortices associated with turbulent fluctuations. Here we analyze the data of direct numerical simulations and study the statistical properties of these motions.

The rest of the paper is organized as follows. In Section~\ref{sec:2}, we discuss the details of direct numerical simulations (DNS). In Section~\ref{sec:3}, we introduce the analytical model of point vortices, which takes into account periodic boundary conditions. Next, in Section~\ref{sec:4}, we present and discuss the obtained numerical results. Finally, we summarize our findings in Section~\ref{sec:5}.

\section{Numerical Methods}\label{sec:2}

We solve the incompressible forced Navier-Stokes equation with linear bottom friction and hyperviscous dissipation for a fluid with unit density in 2D:
\begin{equation}\label{eq:1}
\partial_t \bm v + (\bm v \nabla) \bm v = - \nabla p -\alpha \bm v - \nu (-\nabla^2)^{q} \bm v + \bm f,
\end{equation}
where $\bm v$ is 2D velocity, $p$ is the pressure, $\alpha$ is the friction coefficient, $\nu$ is the hyperviscosity, and $\bm f$ is a random forcing. The domain is a doubly periodic square box of size $L=2 \pi$. The forcing is isotropic in space and shortly correlated in time. Its spatial spectrum is Gaussian with mean $k_f=100$ and standard deviation $\delta_f=1.5$. The short time correlation of the force implies that the average energy injection rate $\varepsilon = \langle \bm v \cdot \bm f \rangle$ is an externally controlled parameter, and here angular brackets denote time and space averaging.

DNS results are obtained by integrating (\ref{eq:1}) in the vorticity formulation using the GeophysicalFlows.jl pseudospectral code~\cite{GeophysicalFlowsJOSS}. The high degree of hyperviscosity $q=8$ allows simulations to be performed at relatively low spatial resolution $512^2$ or $1024^2$, see Refs.~\onlinecite{chertkov2007dynamics, laurie2014universal, frishman2018turbulence, frishman2017jets, parfenyev2022profile}. The time step is fixed for each simulation and it satisfies $\Delta t < c_0 \Delta x/v_{max}$, where $\Delta x$ is the grid spacing, $v_{max}$ is the maximum value of the velocity field projections on the axes of the Cartesian coordinate system, and $c_0$ is equal to $0.3-0.6$ (CFL criterion). The time step $\Delta t$ is also the correlation time of the exciting force $\bm f$. The velocity field can be uniquely reconstructed from the vorticity field; the zero harmonic is absent due to the flow excitation conditions.

In all our simulations, the bottom friction is small enough, $\alpha \ll \varepsilon^{1/3} L^{-2/3}$, so that the energy, transferred to the domain size by the inverse cascade, is accumulated there, giving rise to a coherent flow~\cite{boffetta2012two}. The initial condition is a state of rest, and each simulation is run until the system reaches a non-equilibrium stationary state, observed by the saturation of the total kinetic energy. Due to the utilization of hyperviscosity, some of the injected energy $\varepsilon$ is dissipated at high wave numbers $k>k_f$~\cite{frishman2017jets, frishman2018turbulence, parfenyev2022profile}. To estimate the inverse energy flux, we compute the energy dissipation rate by bottom friction during the steady-state regime, $\epsilon = 2 \alpha \langle \bm v^2/2 \rangle$. This estimate is justified because the main contribution to the total energy of the system comes from large scales. 
Once stationary, we output data at every turnover time of the coherent flow, $\tau_v \sim \sqrt{\alpha L^2/\epsilon}$. Statistics are collected over about $10^5$ turnover times.

Figure~\ref{fig:1}a shows an example of a condensate steady-state, taking the form of a system-sized vortex dipole. The dipole slowly drifts with time in random directions, and the movements of coherent vortices are strongly correlated. Figure~\ref{fig:1}b shows a coherent vortex trajectory unfolded from the computational domain $[-\pi, \pi)^2$ into $\mathbb{R}^2$. We determine the positions $\bm R_+$ and $\bm R_-$ of the vortices by their centers, identified by the maximum and minimum of the vorticity~\cite{frishman2018turbulence, parfenyev2022profile}. The velocities $\bm V_c^+$ and $\bm V_c^-$ of the positive and negative coherent vortices are measured as the ratio of the displacements of the vortex centers on adjacent snapshots to the time between them. The velocities $\bm V_c^{\pm}$ defined in this way refer to the point in time midway between adjacent snapshots.

An analysis of the displacement $\Delta \bm R_+$ of the positive vortex over time $t$ is shown in Fig.~\ref{fig:1}c. One can see a clear transition from ballistic $\langle \Delta \bm R_+^2 \rangle \propto t^2$ to diffusion $\langle \Delta \bm R_+^2 \rangle \propto t$ regime, which is typical for all simulations. This behavior means that the velocities of the vortices have a finite correlation time. Diffusion behavior at larger times is universal due to the central limit theorem. 

\begin{figure*}[t]
\centering{\includegraphics[width=\linewidth]{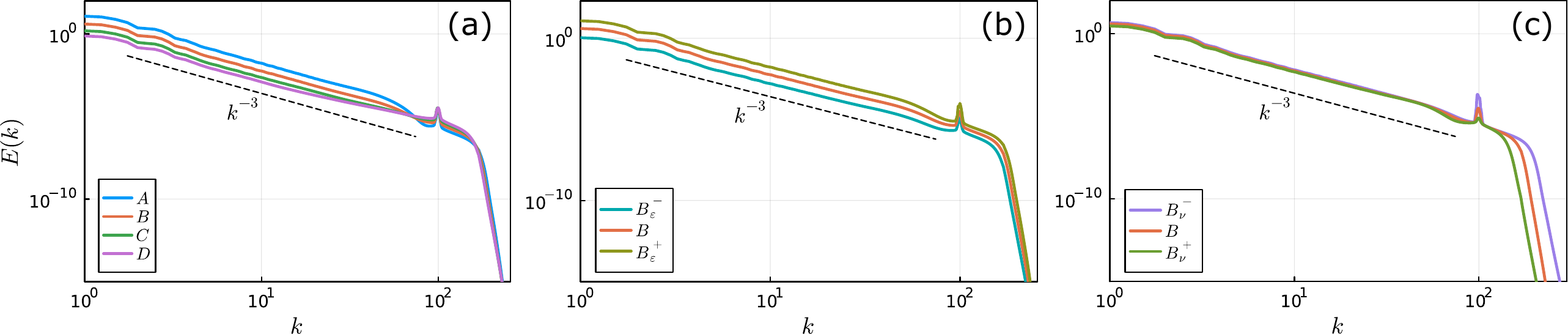}}
\caption{Energy spectra for DNS runs $A$--$D$ (a), runs $B_{\varepsilon}^-$--$B_{\varepsilon}^+$ (b), and runs $B_{\nu}^-$--$B_{\nu}^+$ (c).}
\label{fig:2}
\end{figure*}

\section{Model of Point Vortices}\label{sec:3}

The coherent vortices occupy a small part of the system area, see Fig.~\ref{fig:1}a. Therefore, in the leading approximation, they can be considered as point objects. Let us denote their circulation $\pm G$, and let them be located at the points with coordinates $(x_1, y_1)$ and $(x_2,y_2)$. The vortices induce a velocity field around them and thereby move each other. The corresponding motion can be described in terms of the Euler equation, and its solution for $2a \times 2b$ periodic rectangular domain is known and it can be expressed in terms of the Weierstrass $\zeta$-function. Its definition and main properties are given in Appendix~\ref{app:a}.

Let $z = x + iy$ and then for the velocity field $\mathcal{V}(z)=v_x-i v_y$ at any point of the domain we have~\cite{o1989hamiltonian, stremler1999motion, stremler2010relative}
\begin{equation}\label{eq:PVM}
\mathcal{V}(z) = \dfrac{G}{2 \pi i} \zeta(z-z_1|a,ib) - \dfrac{G}{2 \pi i} \zeta(z-z_2|a,ib) + \Delta,
\end{equation}
where the constant $\Delta$ ensures that there is no zero harmonic in the velocity field, and therefore it matches the analytical expression with the results of DNS, 
\begin{equation}
\Delta = - \dfrac{G}{2\pi i} \dfrac{1}{ab} \left[ \zeta(a|a,ib)b(x_2-x_1) + \zeta(ib|a,ib) a (y_2-y_1) \right].
\end{equation}
Since the Weierstrass $\zeta$-function is anti-symmetric, $\zeta(-z)=-\zeta(z)$, the vortices move with the same constant velocity
\begin{equation}\label{eq:vel}
\dot z_1^* = \dfrac{G}{2\pi i} \zeta(z_2-z_1|a,ib) + \Delta = \dot z_2^*,
\end{equation}
which we call the dipole velocity and denote by $\bm V_d$. It is determined by the mutual arrangement of vortices, which is the integral of motion in the considered model. Moreover, if $z_2-z_1 = a+ib$, then the vortices are immobile, $\bm V_d=0$.

\begin{figure*}[t]
\centering{\includegraphics[width=\linewidth]{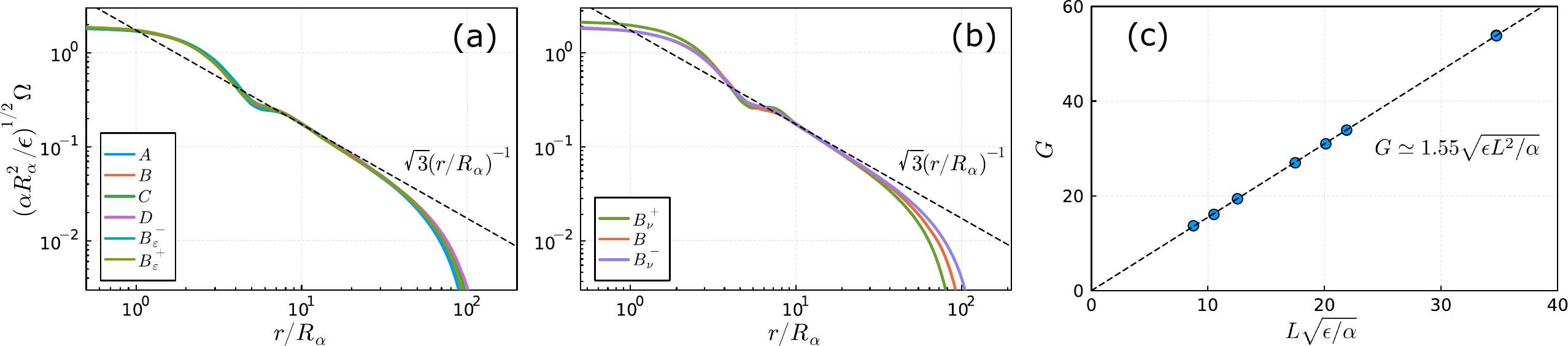}}
\caption{Radial profiles of the mean vorticity for DNS runs $A$--$B_{\varepsilon}^+$ (a) and runs $B_{\nu}^-$--$B_{\nu}^+$ (b). The dashed line corresponds to the analytical result $\Omega(r) = (3\epsilon/\alpha)^{1/2}r^{-1}$. (c) The dependence of the vortex circulation $G$ on the parameter $L \sqrt{\epsilon/\alpha}$. The dashed line corresponds to $G \simeq 1.55 L \sqrt{\epsilon/\alpha}$.}
\label{fig:3}
\end{figure*}

It turned out that in our numerics the distance $z_2-z_1$ between coherent vortices is close to half the diagonal of the computational domain $a+ib$, so expression (\ref{eq:vel}) for the dipole velocity can be simplified by expanding the Weierstrass $\zeta$-function into a Taylor series. It is convenient to choose the coordinate system so that the position of the second vortex (with negative circulation) is at the point $z_2 = a+ib$, and then, using the periodic boundary conditions, we choose the elementary domain so that the position of the first vortex is close to zero, at the point with the coordinate $z_1=\rho_x+i \rho_y$. The real values $(\rho_x, \rho_y)$ uniquely determine the mutual arrangement of the vortices and allow us to determine the dipole velocity $\bm V_d$. For a square domain with $a=b=\pi$, we find
\begin{eqnarray}
\label{eq:d1}
&V_{d}^x \approx -\dfrac{G}{8 \pi^2} \rho_y -\dfrac{G A}{2 \pi} \rho_y (3 \rho_x^2- \rho_y^2),&\\
\label{eq:d2}
&V_{d}^y \approx \dfrac{G}{8 \pi^2} \rho_x -\dfrac{G A}{2 \pi} \rho_x (\rho_x^2 - 3 \rho_y^2),&
\end{eqnarray}
where the constant $A = \Gamma(1/4)^8/3072 \pi^6 \approx 0.01$.

\begin{table}[b]
\caption{\label{tab:1}Parameters for the DNS runs. The second column shows the time step $\tau_v$ of the data output and the last two columns show \textit{a posteriori} estimates of the vortex circulation $G$ and the inverse energy flux $\epsilon = 2 \alpha \langle \bm v^2/2 \rangle$.}
\begin{ruledtabular}
\begin{tabular}{cccccccc}
         run   & grid & $\tau_v$ &  $10^4 \times \varepsilon$  &  $10^4 \times \alpha$ & $10^{35} \times \nu$ & $G$ & $10^4 \times \epsilon$ \\ \hline
         $A$ & $512$   &  $1$ &  $3.5$  &  $0.07$ & $5$ & $53.9$ & $2.14$  \\
         $B$ & $512$    &  $2$ &  $3.5$  &  $0.2$ & $5$ & $31.0$ & $2.05$  \\
         $C$ & $512$    &  $3$ &  $3.5$  &  $0.5$  & $5$ & $19.4$ & $1.99$ \\
         $D$ & $512$    &  $5$ &  $3.5$  &  $1$  & $5$ & $13.7$ & $1.94$ \\
         $B_{\varepsilon}^{-}$ & $512$ & $3$ & $1$ & $0.2$ & $5$ & $16.1$ & $0.56$ \\
         $B_{\varepsilon}^{+}$ & $512$ & $1$ & $10$ & $0.2$ & $5$ & $53.8$ & $6.12$ \\
         $B_{\nu}^{-}$ & $1024$ & $2$ & $3.5$ & $0.2$ & $0.5$ & $33.9$ & $2.43$ \\
         $B_{\nu}^{+}$ & $512$ & $2$ & $3.5$ & $0.2$ & $50$ & $27.0$ & $1.55$ \\
\end{tabular}
\end{ruledtabular}
\end{table}

Since there are turbulent fluctuations, the velocities of coherent vortices are not the same, $\bm V_c^+ \neq \bm V_c^- \neq \bm V_d$. Accordingly, the distance between the vortices changes, and the dipole velocity $\bm V_d$ depends on time. At each moment, we can determine $(\rho_x,\rho_y)$ from the DNS data and calculate $\bm V_d$ using expressions (\ref{eq:d1})--(\ref{eq:d2}). This contribution to the velocity of vortex centers is related to their direct influence on each other. The remaining relatively small part is associated with turbulent pulsations, $\bm V_t^{\pm} \equiv \bm V_c^{\pm} - \bm V_d$. We use linear interpolation to find the dipole velocity $\bm V_d$ between snapshots, where $\bm V_c^{\pm}$ is defined. This is justified because, as will be shown below, the correlation time of $\bm V_d$ is much longer than the time $\tau_v$ between adjacent snapshots (see also Fig.~\ref{fig:1}c). Note also that the values of $\bm V_t^{\pm}$ characterize the displacements of vortex centers over time $\tau_v$ and should not be confused with the instantaneous contributions to the velocities of coherent vortices due to turbulent fluctuations.

\begin{figure*}[t]
\centering{\includegraphics[width=\linewidth]{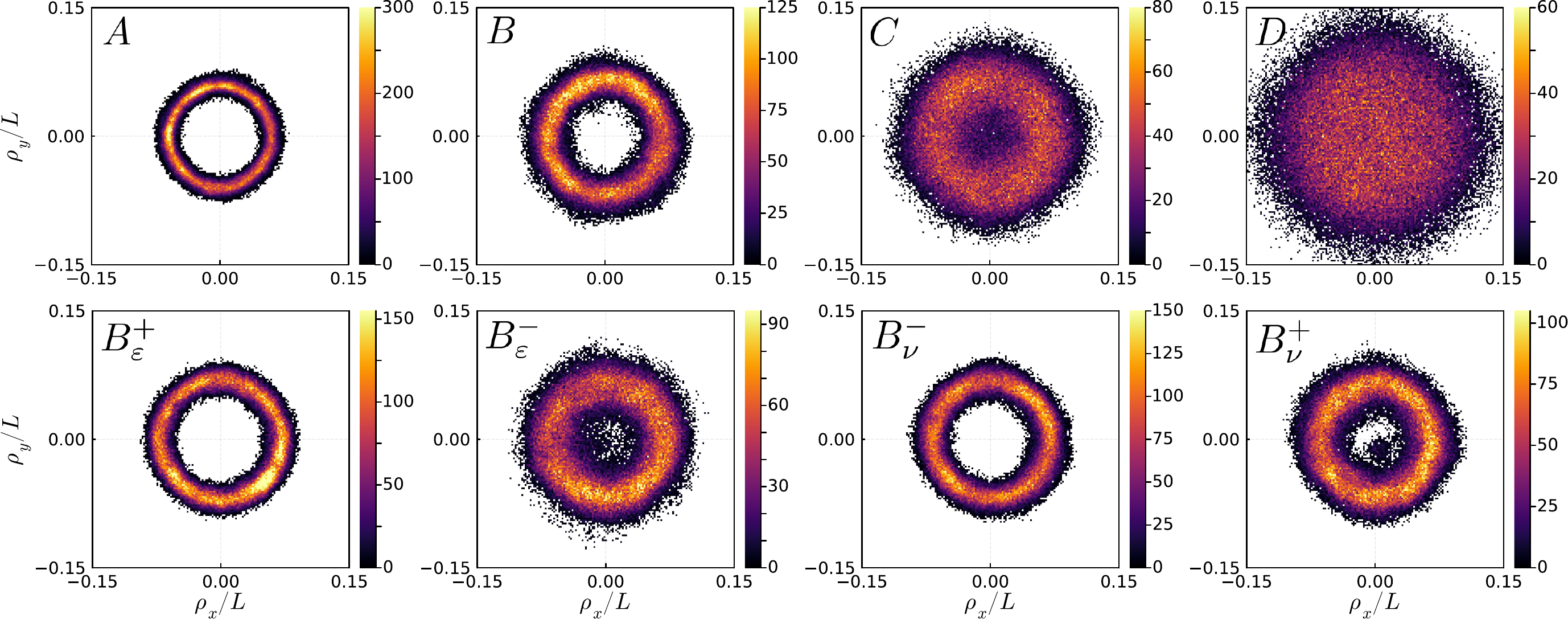}}
\caption{Probability density function  $\mathcal{P}_{2D} (\bm \rho)$ of the vector $\bm \rho$ describing the difference in the mutual arrangement of coherent vortices from half the diagonal of the computational domain.}
\label{fig:4}
\end{figure*}

\section{Results}\label{sec:4}

In this section, we present the results of DNS performed for various values of bottom friction $\alpha$, pumping intensity $\varepsilon$, and fluid hyperviscosity $\nu$. The simulations can be grouped into three sets. In the first set of simulations ($A$--$D$), we change the bottom friction coefficient, while the other parameters are fixed. In the second set of simulations ($B_{\varepsilon}^-$--$B_{\varepsilon}^+$), we vary the pumping intensity, and in the last set of simulations ($B_{\nu}^-$--$B_{\nu}^+$), we change the fluid hyperviscosity. The parameters for the DNS runs are summarized in Table~\ref{tab:1}. In all runs, the Reynolds number at the forcing scale is large, $Re = \epsilon^{1/3}/(\nu k_f^{46/3}) \gg 1$, and the inverse energy cascade reaches the system size, $\delta = \epsilon^{-1/3} L^{2/3} \alpha \ll 1$. Under these conditions, the energy is accumulated at the system size and a coherent vortex dipole is formed.

\begin{figure*}[t]
\centering{\includegraphics[width=\linewidth]{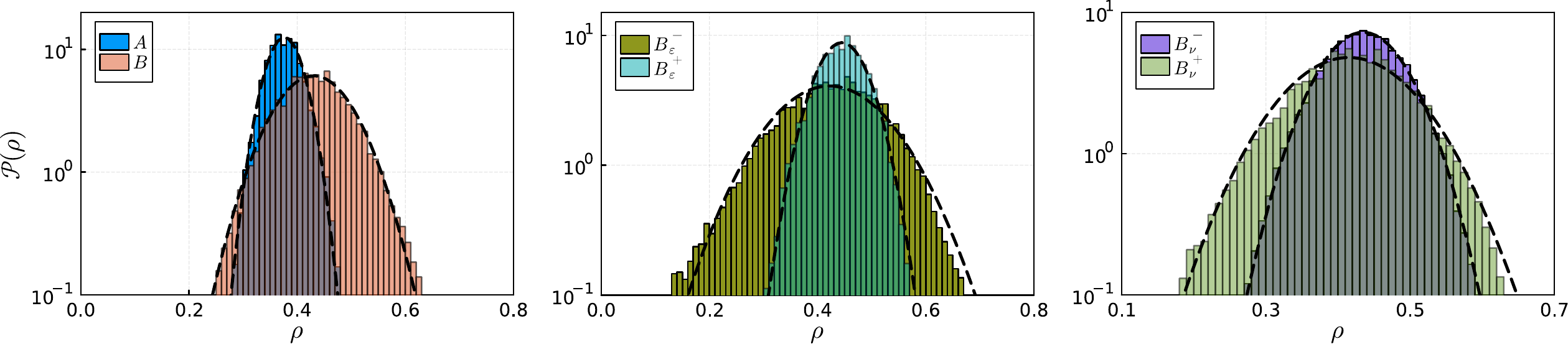}}
\caption{Probability density function $\mathcal{P} (\rho)$ of the modulus of the vector $\bm \rho$. The dashed lines show the approximation $\mathcal{P} (\rho) \propto \rho \exp \left[ -(\rho-\rho_0)^2/2w^2 \right]$ with the parameters given in Table~\ref{tab:2}.}
\label{fig:5}
\end{figure*}

The energy spectra are shown in Fig.~\ref{fig:2}, and they are steeper than $-5/3$ in the region $k<k_f$ due to the presence of the condensate~\cite{chertkov2007dynamics, chan2012dynamics, frishman2017jets, parfenyev2022profile, scott2023annular}. Most of the energy is concentrated at large scales that justifies our estimate $\epsilon = 2 \alpha \langle \bm v^2 \rangle/2$ for the inverse energy flux. Coherent vortices become stronger as bottom friction $\alpha$ decreases. Strong coherent vortices maintain their existence by suppressing turbulent fluctuations at the pumping scale $k_f$, which leads to a deformation of the spectrum in this region, see Fig.~\ref{fig:2}a. Note also that the value of $\epsilon$ increases slightly as bottom friction $\alpha$ decreases, see Table~\ref{tab:1}. An increase in the pumping intensity $\varepsilon$ causes an increase in the spectral energy density $E(k)$ at all scales, see Fig.~\ref{fig:2}b. The fluid hyperviscosity mainly affects the spectral energy density $E(k)$ in the region $k \gtrsim k_f$, see Fig.~\ref{fig:2}c. In the last set of simulations ($B_{\nu}^-$--$B_{\nu}^+$), the parameter $\epsilon$ changes by a factor of $1.5$, see Table~\ref{tab:1}.

The mean vorticity profiles of the coherent vortices in dimensionless units are shown in Figs.~\ref{fig:3}a and \ref{fig:3}b. We normalize velocities by $\sqrt{\epsilon/\alpha}$ and measure distances in units of $R_{\alpha}=(\nu/\alpha)^{1/2q}$, which corresponds to the size of the viscous vortex core~\cite{parfenyev2022profile}. A high degree $q=8$ of hyperviscosity means that $R_{\alpha}$ weakly depends on the problem parameters. The vorticity profiles almost overlap and agree well with the analytical prediction $\Omega(r) = (3\epsilon/\alpha)^{1/2}r^{-1}$ valid outside the viscous core, see Refs.~\onlinecite{laurie2014universal, kolokolov2016structure, frishman2017culmination}. Note that as the fluid hyperviscosity increases, the vorticity profiles become slightly steeper, see Fig.~\ref{fig:3}b, in agreement with the results reported earlier in Ref.~\onlinecite{parfenyev2022profile}. We use the mean vorticity profiles to calculate the vortex circulations $G=\int 2 \pi r \Omega(r) dr$ that enter into expressions (\ref{eq:d1})--(\ref{eq:d2}), see Table~\ref{tab:1}. Substituting the analytical expression for the vortex profile, one can find that $G = 2 \pi R \sqrt{3 \epsilon/\alpha}$, where $R$ is the vortex radius. From the analysis of numerical results, we obtain $R \simeq L/7$ and, respectively, $G \simeq 1.55 L \sqrt{\epsilon/\alpha}$ for all DNS runs, see Fig.~\ref{fig:3}c.

\begin{table}[b]
\caption{\label{tab:2} Dimensionless parameters $\delta=\epsilon^{-1/3} L^{2/3} \alpha$ and $Re$ characterizing the state of the system, fitting parameters $\rho_0$ and $w$ for $\mathcal{P}_{2D} (\bm \rho) \propto \exp \left[ -(\rho-\rho_0)^2/2w^2 \right]$, and superdiffusion exponent $\beta$ in the regime of strong condensate.}
\begin{ruledtabular}
\begin{tabular}{ccccccc}
         run   & $\rho_0$ &  $w$ & $\beta$ & $10^3 \times \delta$ & $Re$ \\ \hline
         $A$     &  $0.37$ &  $0.032$ & $1.85$ & $0.40$ &  $258$   \\
         $B$     &  $0.42$ &  $0.066$ & $1.70$ & $1.15$ &  $254$   \\
         $B_{\varepsilon}^{-}$ & $0.40$ & $0.101$ & $1.55$ & $1.78$ &  $165$ \\
         $B_{\varepsilon}^{+}$ & $0.44$ & $0.046$ & $1.75$ & $0.80$ &  $366$ \\
         $B_{\nu}^{-}$ & $0.43$ & $0.057$ & $1.70$ & $1.09$ &  $2689$ \\
         $B_{\nu}^{+}$ & $0.40$ & $0.085$ & $1.70$ & $1.27$ &  $23$ \\
\end{tabular}
\end{ruledtabular}
\end{table}

The power-law behaviour of mean vorticity profiles $\Omega(r) \propto 1/r$ makes it possible to explain the dependence $E(k) \propto k^{-3}$ for the energy spectrum observed in Fig.~\ref{fig:2}. By performing a two-dimensional Fourier transform, one can establish that $\tilde \Omega(k) \propto 1/k$ and therefore $|\tilde v(k)|^2 \propto |\tilde \Omega(k)|^2/k^2 \propto 1/k^4$. After integration over the rings in $k$-space, we obtain $E(k) \propto k^{-3}$. Note also that if we subtract the coherent part of the flow corresponding to the coherent vortices from snapshots of vorticity field, then the scaling steeper than $k^{-5/3}$ disappears in the energy spectrum for fluctuations in the region $k<k_f$ (not shown).

The mutual arrangement of coherent vortices can be characterized by the two-dimensional vector $\bm \rho = (\rho_x, \rho_y)$, which was introduced in Section~\ref{sec:3} and specifies the difference in the arrangement of vortices from half the diagonal of the computational domain. Its probability density function (PDF) $\mathcal{P}_{2D} (\bm \rho)$ is shown in Fig.~\ref{fig:4}. Remarkably, at small $\delta \lesssim 2 \times 10^{-3}$, the coherent vortices avoid the configuration with $\bm \rho = 0$ (corresponding to $\bm V_d = 0$), and PDF has the form of a ring. We call this regime strong condensate and our study is mainly focused on it. An increase in the dimensionless parameter $\delta$ entails stronger fluctuations in the distance between coherent vortices and leads to smearing of the ring. We found that in the limit of strong condensate, the PDF $\mathcal{P}_{2D} (\bm \rho) $ can be well approximated by
\begin{equation}\label{eq:ring}
\mathcal{P}_{2D} (\bm \rho) \propto \exp \left[- \dfrac{(\rho-\rho_0)^2}{2w^2} \right],
\end{equation}
where $\rho = |\bm \rho|$, and parameters $\rho_0$ and $w$ characterize the radius and width of the ring, correspondingly. The quality of the fit is illustrated in Fig.~\ref{fig:5} for the one-dimensional PDF $\mathcal{P} (\rho) = \int d \psi \rho \mathcal{P}_{2D} (\bm \rho)$, where integration is performed over the polar angle $\psi$. The parameters of the fit and the values of dimensionless parameters $\delta = \epsilon^{-1/3} L^{2/3} \alpha$ and $Re = \epsilon^{1/3}/(\nu k_f^{46/3})$, characterizing the state of the system, are summarized in Table~\ref{tab:2}.

\begin{figure}[t]
\centering{\includegraphics[width=0.8\linewidth]{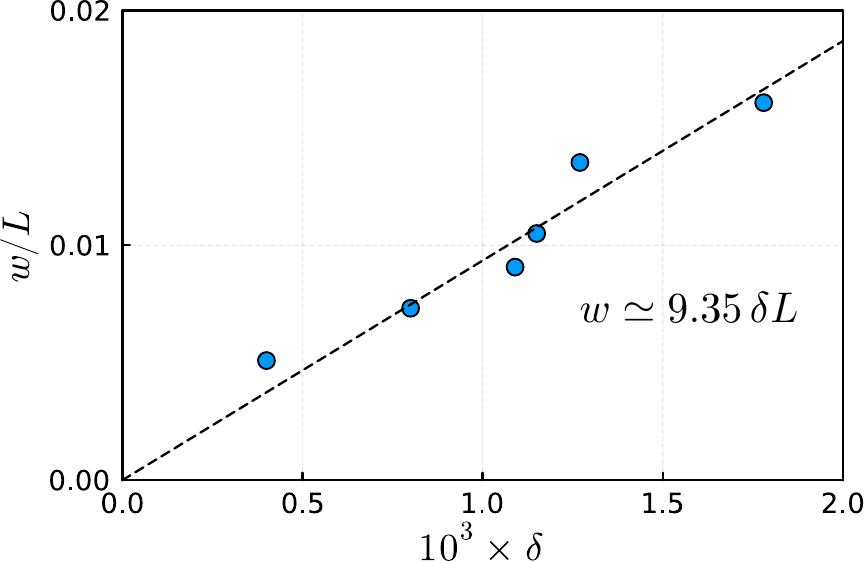}}
\caption{Scaling of the ring width $w$ with the parameter $\delta=\epsilon^{-1/3} L^{2/3} \alpha$. The dashed line corresponds to $w \simeq 9.35 \, \delta L$.}
\label{fig:6}
\end{figure}

\begin{figure*}[t]
\centering{\includegraphics[width=\linewidth]{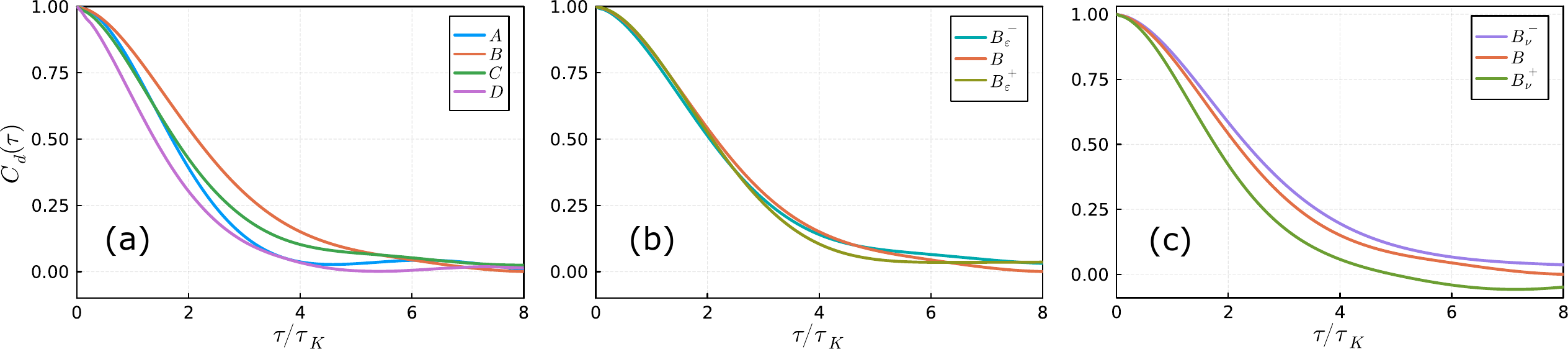}}
\caption{Pair correlation function $C_d(\tau)$ of the dipole velocity for DNS runs $A$--$D$ (a), runs $B_{\varepsilon}^{-}$--$B_{\nu}^{+}$, and runs $B_{\nu}^-$--$B_{\nu}^+$ (c).}
\label{fig:7}
\end{figure*}

The radius $\rho_0$ of the ring weakly depends on control parameters. In our numerics, the relative change in the ring radius did not exceed $20\%$, see Table~\ref{tab:2}. We can note a trend toward a decrease in the ring radius with a decrease in bottom friction $\alpha$, pumping intensity $\varepsilon$, and with an increase in fluid hyperviscosity $\nu$. The width $w$ of the ring, on the contrary, varies over a wide range. Its value is mainly determined by the dimensionless parameter $\delta$. Summarizing the data in Table~\ref{tab:2}, we obtain $w/L \simeq 9.35 \, \delta$, see Fig.~\ref{fig:6}. 

The PDF of the dipole velocity $\bm V_d$ can be recovered from $\mathcal{P}_{2D} (\bm \rho)$ using expressions (\ref{eq:d1})--(\ref{eq:d2}). The first terms in these expressions are dominant, so the PDF of the dipole velocity inherits an axisymmetric form similar to that of $\mathcal{P}_{2D} (\bm \rho)$ up to rescaling. For the mean square of the dipole velocity we find $\langle V_d^2 \rangle \simeq G^2 \langle \rho^2 \rangle/ 4L^4$ or by substituting the expression for circulation $G$ obtained above, $\alpha \langle V_d^2 \rangle/\epsilon \simeq 0.6 \langle \rho^2 \rangle/ L^2$. In the regime of strong condensate (the width of the ring is small compared to its radius), one can estimate $\langle \rho^2 \rangle \approx \rho_0^2 \ll L^2$, so the characteristic value of the dipole velocity is more than an order of magnitude smaller than the polar velocity $U_{\varphi}=\sqrt{3 \epsilon/\alpha}$ inside coherent vortices. Detailed calculations confirm this simplified consideration. We found $\sqrt{\alpha \langle V_d^2 \rangle/ \epsilon} \simeq$ \{0.047, 0.053, 0.055, 0.063, 0.053, 0.055, 0.054, 0.052\} for DNS runs $A$--$B_{\nu}^{+}$ in the same order as in Table~\ref{tab:1}.

\begin{figure*}[t]
\centering{\includegraphics[width=\linewidth]{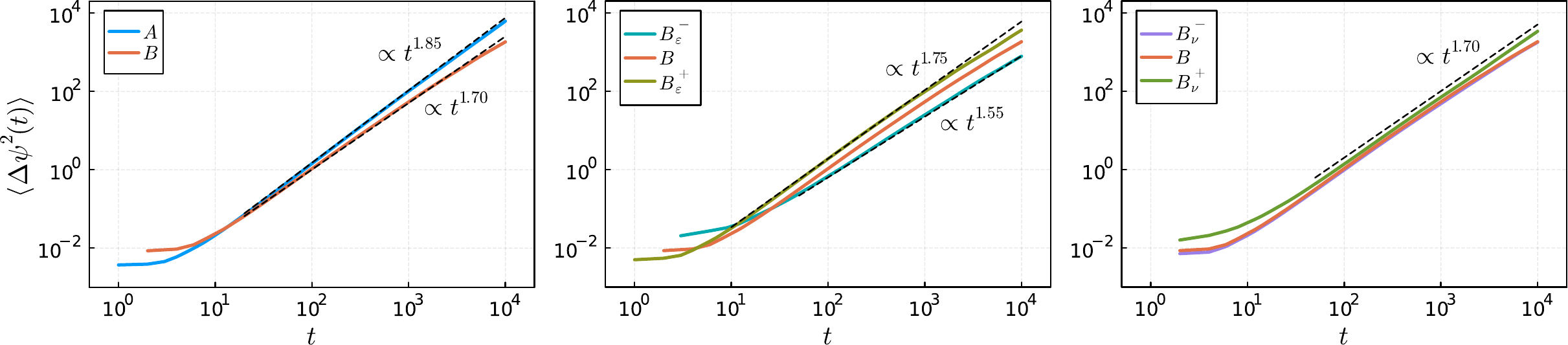}}
\caption{Random walk over the polar angle $\psi$ exhibits superdiffusion behaviour. The data are given for DNS runs corresponding to the strong condensate regime.}
\label{fig:8}
\end{figure*}

Next, we analyze the differences between the velocities of the vortex centers and the dipole velocity, $\bm V_t^{\pm} = \bm V_c^{\pm} - \bm V_d$, which are caused by turbulent fluctuations. We found that their probability distribution functions are Gaussian
\begin{equation}\label{eq:diff}
\mathcal{P}_{2D}(\bm V_t^{\pm}) = \dfrac{1}{2 \pi \sigma^2} \exp \left( -(\bm V_t^{\pm})^2/2\sigma^2 \right),
\end{equation}
with normalized standard deviations $\sqrt{\alpha \sigma^2/\epsilon} \simeq$ \{0.003, 0.004, 0.008, 0.013, 0.007, 0.004, 0.004, 0.005\} for DNS runs $A$--$B_{\nu}^{+}$, respectively. These contributions are small compared to the root mean squared values of the dipole velocities, which means that the model of point vortices correctly describes the main contribution to the velocities of vortex centers. Note that for the regime of strong condensate, the values of the standard deviations $\sigma$ are comparable to the characteristic errors in determining the dipole velocity $GL/(8 \pi^2 n)$ and the velocities of the vortex centers $L/(\tau_v n)$ associated with the finite size ($n=512$ or $n=1024$) of the computational grid. However, this remark does not affect the main conclusion that the turbulent contributions $\bm V_t^{\pm}$ to the motion of coherent vortices are small compared to the dipole velocity $\bm V_d$.

Despite their relative smallness, the contributions $\bm V_t^{\pm}$ to the velocities of vortex centers are important, since they lead to a change in the mutual arrangement of coherent vortices and hence to a finite correlation time of the dipole velocity, see Fig.~\ref{fig:1}c. This process can be thought of as a random walk around the PDF $\mathcal{P}_{2D} (\bm \rho)$ shown in Fig.~\ref{fig:4}. To quantitatively describe the time correlations of the dipole velocity, we introduce the pair correlation function
\begin{equation}\label{eq:PCF}
C_d(\tau) = \langle \bm V_d(t) \bm V_d(t+\tau) \rangle/\langle \bm V_d^2 \rangle,
\end{equation}
where the angular brackets mean averaging over time $t$. It turns out that the correlation time of the dipole velocity $\tau_d$ is long compared to the turnover time of coherent flow, but short compared to the decay time due to bottom friction, $\tau_v \ll \tau_d \ll 1/\alpha$. By order of magnitude, this time is comparable to the turnover time $\tau_K = L^{2/3} \epsilon^{-1/3}$ of system-size eddies produced by an inverse turbulent cascade, see Fig.~\ref{fig:7}. This result is consistent with Ref.~\onlinecite{chan2012dynamics}, where bottom friction $\alpha$ was set to zero and simulations were run with normal viscosity instead of hyperviscosity. When varying pumping intensity, time rescaling $\tau/\tau_K$ allows the pair correlation functions of the dipole velocity to be matched to the master curve, see Fig.~\ref{fig:7}b. 

\begin{figure}[t]
\centering{\includegraphics[width=0.8\linewidth]{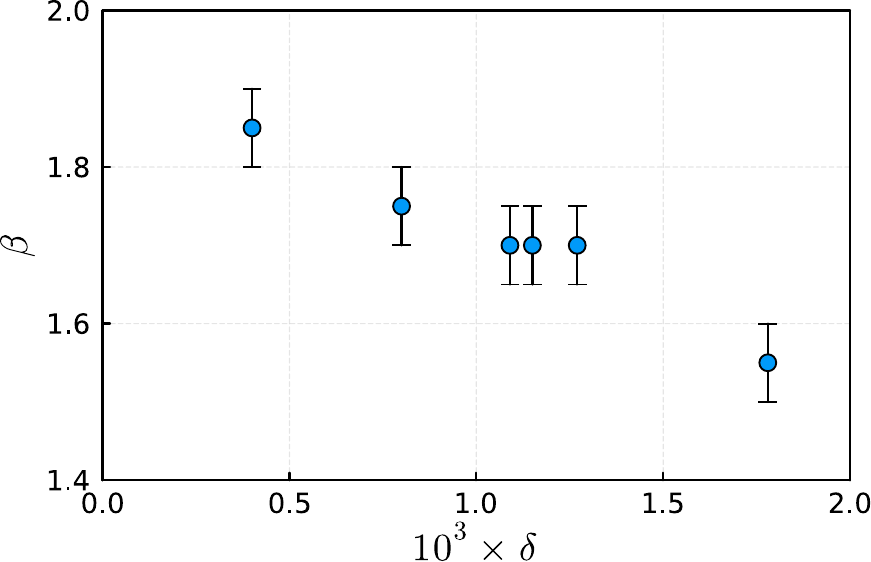}}
\caption{Dependence of superdiffusion exponent $\beta$ on the parameter $\delta$.}
\label{fig:9}
\end{figure}

For a strong condensate, the decrease in $C_d(\tau)$ is mainly related to a random walk over the polar angle $\psi$, see Fig.~\ref{fig:4}. To characterize this random process, we unfold the angle variable from domain $[-\pi, \pi)$ into $\mathbb{R}$, and calculate the mean displacement squared $\langle \Delta \psi^2(t) \rangle$ over time $t$. Surprisingly, we obtain the superdiffusion behavior $\langle \Delta \psi^2(t) \rangle \propto t^{\beta}$ with $1.5<\beta<2$ at times $t \gg \tau_v$, see Fig.~\ref{fig:8}. The exponent $\beta$ grows as the bottom friction $\alpha$ decreases and the pumping intensity $\varepsilon$ increases, and it is practically independent of the fluid hyperviscosity $\nu$. These interdependencies can be simplified to a single parameter $\delta$, see Fig.~\ref{fig:9}. The superdiffusion behavior indicates the existence of long-time correlations in our system that exist over thousands of coherent flow turnover times. The duration of these correlations is also long compared to the correlation time $\tau_d$ of the dipole velocity. The collected statistics do not allow us to determine the behavior of correlation functions at times comparable to $1/\alpha$ and longer.

\section{Conclusion}\label{sec:5}

To summarize, we studied the motion of vortex condensate, which is formed in the two-dimensional periodic square domain, when the Reynolds number at the forcing scale is large, $Re \gg 1$, and when the inverse energy cascade reaches the system size, $\delta = \epsilon^{-1/3} L^{2/3} \alpha \ll 1$. Under these conditions, the energy accumulates at the system size in the form of a vortex dipole, and we mainly focused on the regime of strong condensate, which corresponds to $\delta \lesssim 2 \times 10^{-3}$. We established that coherent vortices can be described as point vortices in the main approximation. Within this model, they move at a constant dipole velocity, which is determined by their circulation and mutual arrangement through expressions (\ref{eq:d1})--(\ref{eq:d2}).

Weak turbulent fluctuations lead to a small relative velocity between the vortices, which changes the distance between them and makes the dipole velocity time-dependent. We found that the distance between coherent vortices is close to half the diagonal of the computational domain, but the vortex dipole avoids being exactly in this configuration, when the parameter $\delta$ is small enough. In this case, the probability density function $\mathcal{P}_{2D} (\bm \rho)$ of the vector $\bm \rho$, describing the difference in the mutual arrangement of coherent vortices from half the diagonal of the computational domain, has the form of a ring. The radius $\rho_0$ of the ring weakly depends on bottom friction $\alpha$, pumping intensity $\varepsilon$, and fluid hyperviscosity $\nu$, while the width $w$ of the ring is controlled by the dimensionless parameter $\delta = \epsilon^{-1/3} L^{2/3} \alpha$, see Fig.~\ref{fig:6}.

Each point on the ring corresponds to a certain arrangement of coherent vortices, and turbulent fluctuations lead to a random walk around the ring, which has superdiffusion behavior $\langle \Delta \psi^2(t) \rangle \propto t^{\beta}$ with $1.5<\beta<2$. The exponent $\beta$ decreases with the parameter $\delta$, see Fig.~\ref{fig:9}. In particular, the random walk results in a finite correlation time of the dipole velocity, which turns out to be of the order of turnover time $\tau_K = L^{2/3} \epsilon^{-1/3}$ of system-size eddies produced by an inverse turbulent cascade, see Fig.~\ref{fig:7}.

The results obtained show a number of features in the behaviour of coherent vortices, the explanation for which is currently lacking. Why do vortices avoid being located exactly on half the diagonal of the computational domain, when the parameter $\delta$ is small enough? What is the nature of the time correlations that exist in the system over thousands of turnover times of coherent vortices, and whose fingerprint was observed in superdiffusion? Answers to these questions will require the development of a theory that describes the effective interaction between coherent vortices via turbulent fluctuations, and we believe that the scaling relations established here will be valuable along this way.

\appendix
\section{Weierstrass $\zeta$-function}\label{app:a}

The Weierstrass $\zeta$-function is defined by
\begin{equation}\label{zeta}
\zeta(z|\varpi,\varrho)= \frac{1}{z}+\sum_{m,m_1}' \left\{\frac{1}{z-\Omega_{mm_1}}+
\frac{1}{\Omega_{mm_1}}+\frac{z}{\Omega_{mm_1}^2}\right\},
\end{equation}
with $\Omega_{mm_1}=2m\varpi+2m_1\varrho$ for integers $m$ and $m_1$, and with a prime indicating that the sum is over all pairs of integers $(m,m_1) \neq (0,0)$. By collecting all the terms in brackets, it can be shown that the series is absolutely convergent. From the above expression, it is clear that the function is antisymmetric, $\zeta(-z)=-\zeta(z)$. Note that each of the series separately in expression (\ref{zeta}) is not convergent. To prove the identities of interest, it is convenient to use symmetric regularization ($-M<m,m_1<M$ and $M \to \infty$), in which the second term becomes zero. In this way, one can show that
\begin{eqnarray}
&\zeta(z+2\varpi|\varpi,\varrho)= \zeta(z|\varpi,\varrho)+2\zeta(\varpi|\varpi,\varrho),&\\
&\zeta(z+2\varrho|\varpi,\varrho)= \zeta(z|\varpi,\varrho)+2\zeta(\varrho|\varpi,\varrho),&\\
&\zeta(\varpi+\varrho|\varpi,\varrho)= \zeta(\varpi|\varpi,\varrho)+\zeta(\varrho|\varpi,\varrho).&
\end{eqnarray}
Note that the Weierstrass $\zeta$-function is not doubly periodic, which leads to the $\Delta$-term in expression (\ref{eq:PVM}).

\acknowledgments

I am grateful to I.V. Kolokolov, V.V. Lebedev and S.S. Vergeles for valuable discussions. This work was performed in the Laboratory ``Modern Hydrodynamics'' created within the framework of Grant 075-15-2022-1099 of the Ministry of Science and Higher Education of the Russian Federation at the Landau Institute for Theoretical Physics and was supported by the Russian Science Foundation (Project No. 23-72-30006) and the Foundation for the Advancement of Theoretical Physics and Mathematics ``BASIS'' (Project No. 22-1-3-24-1). Simulations were performed on the cluster of the Landau Institute.

\section*{Data AVAILABILITY}

The data that support the findings of this study are available from the corresponding author upon reasonable request.

\bibliography{biblio}

\end{document}